\documentclass[pr,notitlepage,superscriptaddress,%nofootinbib,
nobibnotes,twocolumn]{revtex4-2}
\usepackage[english]{babel}
\usepackage{mathtools}
\usepackage{graphicx}
\usepackage{mathrsfs}
\usepackage{amssymb}
\usepackage{xcolor}
\usepackage{amsmath}
\usepackage{bm}
\usepackage{physics}

\usepackage[unicode=true,colorlinks=true,citecolor=blue,urlcolor=blue]{hyperref}

\newcommand{\mysection}[1]{\textcolor{blue}{\textit{#1}.}}

\begin{document}
\title{Hanle effect in current induced spin orientation}

\author{L.~E.~Golub} 
\affiliation{Physics Department, University of Regensburg, 93040 Regensburg, Germany}	
\author{E.~L.~Ivchenko} 	
\affiliation{Ioffe Institute,  	194021 St.~Petersburg, Russia}

\begin{abstract}
Electrical spin orientation is the generation of electron spin proportional to the electric current. This phenomenon is allowed by symmetry in gyrotropic systems, e.g. in inversion-asymmetric structures with Rashba spin-orbit splitting. Here we develop a theory of electrical spin orientation for magnetic two-dimensional heterostructures. Spin-orbit coupled graphene and semiconductor heterostructures proximitized by ferromagnets are considered. The analytical theory is based on the Boltzmann kinetic equation for a spin-dependent distribution function and collision integral. We show that the induced spin demonstrates the Hanle effect: a direction of the spin depends on the out-of-plane magnetization. Importantly, the Hanle effect is extremely sensitive to the details of electron elastic scattering. In semiconductor heterostructures, the effect of magnetization is present for scattering by long-range disorder and absent for short-range scattering. In spin-orbit-coupled graphene, the Hanle effect occurs at any scattering potential, but the direction of the spin strongly changes with variation of the disorder type. The theory also describes the effect of valley-Zeeman splitting on the electrical spin orientation in graphene, where the spin experiences opposite Hanle effects in two valleys.
\end{abstract}

\maketitle

\mysection{Introduction}
Gyrotropic systems, where at least some components of vectors and pseudovectors transform according to equivalent representations of the symmetry point group, allow the current-induced spin orientation or polarization (CISP). This phenomenon consists in a generation of electron spin induced by the electric current as a linear response:
\begin{equation}
\label{CISP_phenom}
\bm s = \hat{\bm Q}\bm j.
\end{equation}
Here $\bm s$ and $\bm j$ are spin and current densities, and $\hat{\bm Q}$ is a second-rank pseudotensor.
The CISP, predicted in semiconductor crystals~\cite{Ivchenko1978,Aronov1989} and semiconductor heterostructures~\cite{Vasko1979,Edelstein1990}, is being investigated nowadays in various systems both theoretically and experimentally, for a recent review see Ref.~\cite{Ganichev2024}.
Mostly it is studied in inversion asymmetric two-dimensional (2D) structures where a common microscopic ground for the effect is the Rashba spin-orbit coupling. Examples are 2D electron gas (2DEG) in semiconductor nanostructures and spin-orbit coupled graphene.
In these systems, the tensor $\hat{\bm Q}$ has one linearly-independent component $Q_{xy} = - Q_{yx} \equiv Q$, so that $\bm s = Q [\bm j \times \hat{\bm z}]$, where $z$ is a direction normal to the 2D plane. This means that the spin is oriented in this plane perpendicularly to the electric current direction.

\begin{figure}[h]
	\centering \includegraphics[width=\linewidth]{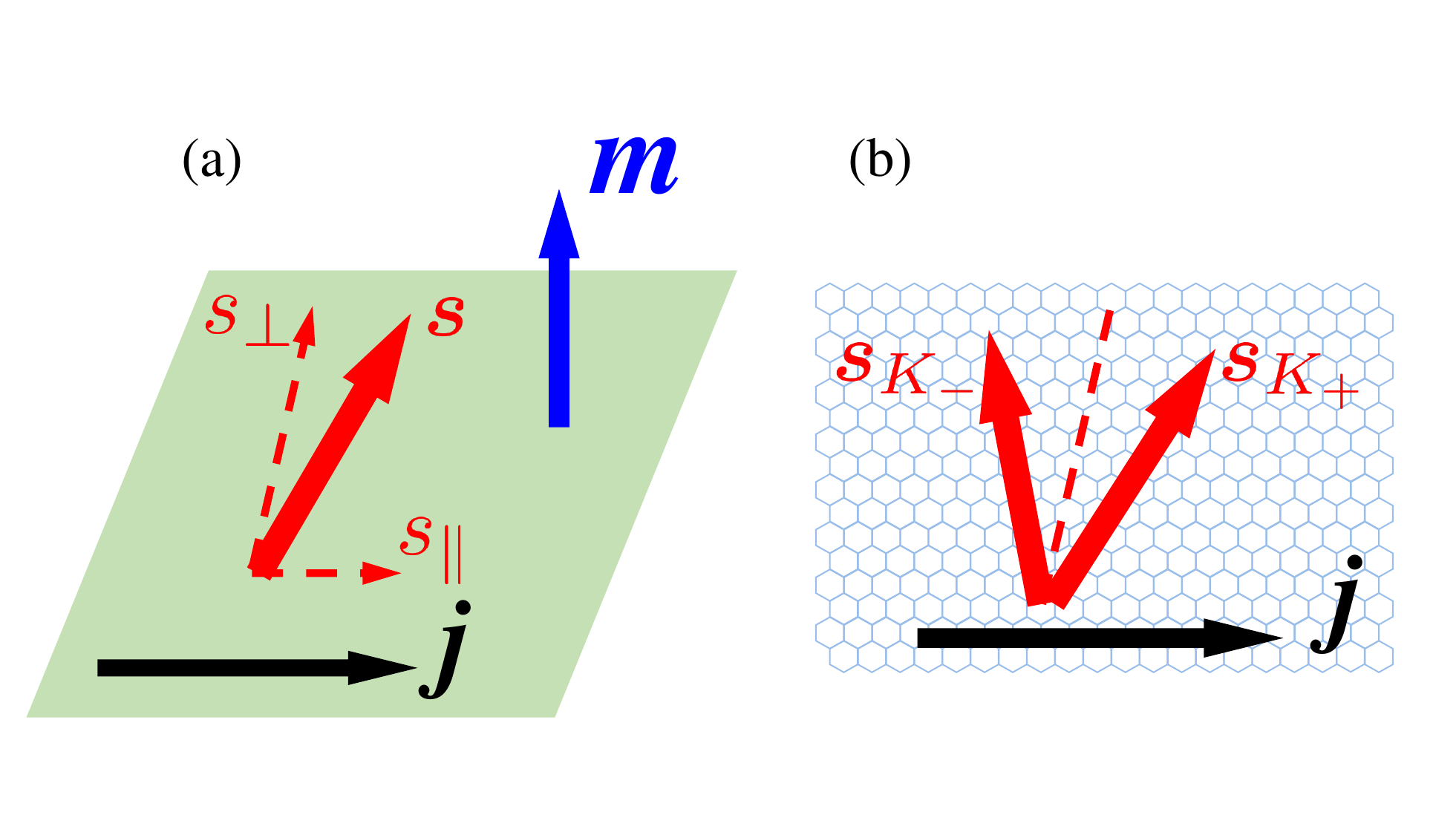} 	
	\caption{Current induced spin orientation in the presence of the Zeeman splitting. In  Rashba 2DEG (a), an out of plane magnetization $\bm m$ results in the Hanle effect: the spin $\bm s$ oriented by the current $\bm j$ has a longitudinal component $s_\parallel$.
% due to the Larmor precession. 
In  spin-orbit coupled graphene (b), the valley-Zeeman splitting results in the spin orientation with a longitudinal spin components opposite in two valleys.}
	\label{Fig_Hanle_CISP}
\end{figure}

A new chapter in studies of the CISP effect opened when an electric current injection was applied to induce a spin torque in the ferromagnetic heterostructure reversing the magnetization of the magnetic layer, see e.g. review~\cite{Gambardella2011}. 
In this case the Zeeman splitting of the electron states in the nonmagnetic layer arises due to proximity effects. 
In the spin-orbit coupled graphene, e.g. in a hybrid graphene-TMDC system, the role of Zeeman field plays the valley-Zeeman (spin-valley) splitting opposite in two valleys~\cite{Garcia2018,Colloq_RMP2020,Sierra2021}. 
Furthermore, in heterostructures containing graphene in close proximity with a  ferromagnetic layer, a magnetization can couple with the Dirac fermion spins \cite{Abtew2013,Wang2015}.
Ingredients of the band model to describe the CISP and spin torque are the Rashba spin-orbit coupling of the free electron spin with its translational motion and the exchange interaction between the conduction electron spin and the ferromagnetic magnetization. The conducting electron effective Hamiltonian has the following form
\begin{equation}
\label{H_general}
\mathcal H  = \varepsilon_k + {\hbar \over 2} \bm \sigma \cdot \bm \Omega_{\bm k} + V(\bm r).
\end{equation}
Here $\bm k$ is quasimomentum, $\bm \sigma$ is a vector of spin Pauli matrices, $\varepsilon_k$ is the 2D energy dispersion without spin-orbit interaction, $V$ is the disorder potential, and the effective Larmor frequency reads
$\bm \Omega_{\bm k}=\bm \Omega_{\bm k}^{\rm R}+\bm \Omega^{\rm Z}$,
where
\begin{equation}
\label{Omega}
\bm \Omega_{\bm k}^{\rm R} =\Omega_{\rm R} [\bm k \times \hat{\bm z}]/k, \quad
\bm \Omega^{\rm Z}= \Omega_{\rm Z} \bm m,
\end{equation}
${\bm m}$ is the magnetization direction of the ferromagnet,
and 
$\Omega_{\rm R,Z}$ are the Rashba and Zeeman precession frequencies.

Both out- and in-plane magnetization effects on CISP were 
studied in
a wide range of 2D systems~\cite{Engel2007,Milletari2008,Manchon2008, Wang2010,Wang2012, OrtizPauyac2013,Garello2013,Chen2017,Dyrdal2014,Dyrdal2015,Dyrdal2017, Ndiaye2017,Shumilin2018,Ghiasi2019,Vaz2019,Sousa2020, Rousseau2021, Veneri2022,Lazrak2024,Pigon2024}.
The out-of-plane Zeeman splitting can lead to the Hanle effect: Electrically induced spin precesses, and the spin component $s_\parallel$ in the direction of the current appears in the steady state,
Fig.~\ref{Fig_Hanle_CISP}. 
In this Letter we develop a kinetic theory of the Hanle effect in CISP.
As compared to previously considered various limiting cases, we assume here arbitrary relations between $\Omega_{\rm R}$, $\Omega_{\rm Z}$ and the transport relaxation rate $\tau^{-1}_{\rm tr}$ as well as arbitrary anisotropy of the elastic scattering probability. 
We fix a direction of the magnetization unit vector ${\bm m}$ along the normal $z$, however
the developed theory can readily be generalized to any directions of ${\bm m}$. 
The theory is applied for both 2DEG with parabolic energy dispersion and $\Omega_{\rm R} \propto k$, and spin-orbit coupled graphene (Gr) with a linear dispersion and $\Omega_{\rm R}$ independent of $k$.

\mysection{Theory}
In the following the distribution of electrons in the wave vector ${\bm k}$ and spin spaces is described by the spin-density matrix $\rho_{\bm k}$ which can be decomposed into a sum $f_{\bm k} + {\bm \sigma} \cdot {\bm S}_{\bm k}$, where $f_{\bm k}$ is the distribution function and ${\bm S}_{\bm k}$ is the electron average spin in the ${\bm k}$ state.

Scattering by disorder potential $V(\bm r)$ is characterized in the kinetic theory by the times $\tau_{n}$ describing elastic relaxation of the $n$th angular harmonics of the distribution function ($n=1,2,\ldots$). 
They are given by 
\begin{equation}
\label{tau_n}
{1\over \tau_n(\varepsilon_k)} = {2\pi\over \hbar}\expval{ \mathcal K_{\bm k'\bm k}(1-\cos{n\theta})}_\theta g(\varepsilon_k),
\end{equation}
where $g(\varepsilon_k)$ is the density of states, $\theta=\varphi_{\bm k'}-\varphi_{\bm k}$ is the scattering angle
with $\varphi_{\bm k}$ being the azimuth angle of $\bm k$,
and angular averaging is performed at $k'=k$.
Here 
\begin{equation}
\label{Corr}
\mathcal K_{\bm k'\bm k}^{\rm 2DEG}= \mathcal K_0(q), \quad \mathcal K_{\bm k'\bm k}^{\rm Gr}= \mathcal K_0(q){1+\cos\theta\over 2}, 
\end{equation}
where 
$q=\abs{\bm k' - \bm k}$, and 
$\mathcal K_0(q)$ is the Fourier-image of the scattering potential correlator $\mathcal K_0(\abs{\bm r-\bm r'})=\expval{V(\bm r)V(\bm r')}$.
We remind that the transport relaxation time $\tau_{\rm tr}=\tau_1(\varepsilon_{\rm F})$, where $\varepsilon_{\rm F}$ is the Fermi energy.

The kinetic equation for the electron spin density matrix in the presence of the electric field $\bm E$ reads 
$(e\bm E\cdot \bm \nabla_{\bm k}\rho_{\bm k}+i[\mathcal H_{\bm k}, \rho_{\bm k} ])/\hbar={\rm St}[\rho]$,
where the collision integral is given by~\cite{Ivchenko1990,Shytov2006,Golub2011}
\begin{equation} \label{Coll}
{\rm St}[\rho]
= \frac{2\pi}{\hbar} \sum_{{\bm k}'}
{\cal K}_{{\bm k}'{\bm k}} \left\{ \rho_{{\bm k}'} - \rho_{\bm k}, \delta (\varepsilon_k + \mathcal H_{\bm k} - \varepsilon_{k'} - \mathcal H_{{\bm k}'} ) \right\},
\end{equation}
$\mathcal H_{\bm k} = (\hbar/2) {\bm \Omega}_{\bm k}\cdot {\bm \sigma}$, the curly brackets in $\{ A,B\}$ mean a symmetrized product $(AB+BA)/2$, and the $\delta$-function is understood as $\delta (\varepsilon_{k}  - \varepsilon_{k'} ) + (\mathcal H_{\bm k} - \mathcal H_{{\bm k}'})\partial_{\varepsilon_k} \delta (\varepsilon_{k}  - \varepsilon_{k'})$.
Hereafter we assume the energy relaxation to be slower than the Dyakonov--Perel spin relaxation.
Then the steady-state density matrix at each energy
is found from the equation with the elastic collision integral~\eqref{Coll} only.

Kinetics of the electron spin density $\bm S_{\bm k}$ 
is described by the following equation
\begin{equation}
\label{spin_dyn}
\bm S_{\bm k} \times \bm \Omega_{\bm k} =
\frac{2\pi}{\hbar} \sum_{{\bm k}'}
{\cal K}_{{\bm k}'{\bm k}} \left( {\bm S}_{{\bm k}'} - {\bm S}_{\bm k} \right)  \delta (\varepsilon_k  - \varepsilon_{k'} ) + \bm G_{\bm k}.
\end{equation}
Here 
$\bm G_{\bm k}$ is the generation rate linear in the electric field $\bm E$.
General expression for the generation rate has the form~\cite{Golub2011}
\begin{multline}
\label{G_general}
\bm G_{\bm k} = -{e\over 2}
(\bm E \cdot \bm \nabla_{\bm k})
\qty[f_0'(\varepsilon_k)\bm \Omega_{\bm k}]
\\ - \pi \sum_{\bm k'} 
\mathcal K_{\bm k'\bm k} (\delta f_{\bm k}-\delta f_{\bm k'})(\bm \Omega_{\bm k} - \bm \Omega_{\bm k'}) \partial_{\varepsilon_k} \delta (\varepsilon_{k}  - \varepsilon_{k'}),
\end{multline}
where $\delta f_{\bm k}=-e \tau_1(\varepsilon_k) f_0'(\varepsilon_k) (\bm E \cdot \bm v_{\bm k})$ 
is the standard spin-independent correction to the distribution function with 
$f_0(\varepsilon)$ being the Fermi-Dirac distribution function,
the prime means differentiation over $\varepsilon_k$, and $\bm v_{\bm k}=\hbar^{-1}\bm \nabla_{\bm k} \varepsilon_k$ is the electron velocity.

The in-plane spin generation rate 
is given by
\begin{equation}
\label{G_par}
\bm G_{\bm k}^{\rm in} = 
-e F(\varepsilon_k)
{\hbar \over 2} \qty[\bm \Omega_{\bm k}^{\rm R}(\bm E \cdot  \bm v_{\bm k})- \overline{\bm \Omega_{\bm k}^{\rm R} (\bm E \cdot \bm v_{\bm k})}],
\end{equation}
where $F(\varepsilon_k)$ is different for 2DEG and graphene and given below.
Hereafter the bar denotes averaging over directions of $\bm k$ at fixed energy.
The out of plane component $G_{\bm k, z}$ is unaffected by the scattering details, because the Zeeman splitting is momentum-independent ($\Omega_{\bm k,z} -  \Omega_{\bm k',z}=0$) and the second term in Eq.~\eqref{G_general} vanishes.
As a result, one has
\begin{equation}
\label{G_z}
G_{\bm k, z} = -e f_0''(\varepsilon_k) {\hbar \over 2}\Omega_{\rm Z}(\bm E \cdot \bm v_{\bm k}).
\end{equation}

The solution of Eq.~\eqref{spin_dyn} is found in Appendix~\ref{App_spin_dyn}.
Its angular-averaged part $\overline{\bm S}$
reads
\begin{multline}
\label{S_plus}
\overline{S}_\perp+ i \overline{S}_\parallel=-{\hbar \Omega_{\rm R} eEv_k\over 4} 
\\ \times {(\Omega_{\rm R}^2/ 2) \tau_2 F(\varepsilon_k) +\qty(\Omega_{\rm Z}^2\tau_2-i\Omega_{\rm Z})f_0''(\varepsilon_k) \over
{\Omega_{\rm R}^2/ 2}+(\Omega_{\rm Z}^2\tau_2-i\Omega_{\rm Z})/\tau_1},
\end{multline}
where the subscripts $\parallel, \perp$ denote projections onto  $\bm E$ and $\bm E\times \hat{\bm z}$ directions in the 2D plane.
This expression is valid for any relation between $\Omega_{\rm R}$, $\Omega_{\rm Z}$ and $1/\tau_{1,2}$.

We calculate the electron spin polarization $\bm s= \sum_{\bm k}\bm S_{\bm k}/N$, where $N$ is the electron concentration. Below we show that the result has the following form
\begin{equation}
\label{s_result}
s_\perp + is_\parallel=s_\perp(0)
\qty[1+\eta{\Omega_{\rm Z}\tau_s(\Omega_{\rm Z}\tau_2-i)\over1+\Omega_{\rm Z}\tau_s(\Omega_{\rm Z}\tau_2-i)}],
\end{equation}
where all values 
should be taken at $\varepsilon_k=\varepsilon_{\rm F}$, and we introduced the  characteristic time $\tau_s$ defined by
\begin{equation}
{1\over \tau_s}={\Omega_{\rm R}^2\tau_{\rm tr}\over 2},
\end{equation}
which at $\Omega_{\rm R}\tau_{\rm tr} \ll 1$ becomes the in-plane spin relaxation time.
The factor $\eta$ differs for Rashba 2DEG and spin-orbit coupled graphene and depends on the scattering mechanism.

\mysection{2DEG with Rashba spin-orbit coupling}
In 2DEG with parabolic energy dispersion $\varepsilon_k=\hbar^2k^2/(2m^*)$, we have $\bm v_{\bm k}=\hbar \bm k/m^*$, a constant density of  states,
 the linear-in-$k$ Rashba splitting $\hbar\Omega_{\rm R}=2\alpha k$, and ${1/ \tau_s} = 2\qty({\alpha k_{\rm F}/ \hbar})^2 \tau_{\rm tr}$, where $\alpha$ is the Rashba constant.
According to Ref.~\cite{Golub2011}, the function $F(\varepsilon_k)$ in Eq.~\eqref{G_par} for 2DEG has the form
\begin{equation}
\label{F_2DEG}
F={(\tau_1 f_0')'\over \tau_2},
\end{equation}
where the relaxation times $\tau_{1,2}$ are introduced in Eq.~\eqref{tau_n}.

Substituting $F$ from Eq.~\eqref{F_2DEG} into Eq.~\eqref{S_plus}
and integrating over $\varepsilon_k$ by parts by taking into account that $\Omega_{\rm R}^2 \propto \varepsilon_k$, we obtain Eq.~\eqref{s_result} where
\begin{equation}
\label{eta_2DEG}
s_\perp(0)=-{\alpha e E\tau_{\rm tr} \over  2\hbar\varepsilon_{\rm F}}=-{\alpha jm^* \over 2\hbar\varepsilon_{\rm F}eN},
\quad
\eta_{\rm 2DEG}={\dd \ln\tau_{\rm tr}\over \dd \ln\varepsilon_{\rm F}},
\end{equation}
with $j$ being the electric current density, see Eq.~\eqref{CISP_phenom}.

The important particular case is the axially symmetric scattering by short-range potentials $V({\bm r})$ with a delta-functional correlator $\mathcal K_0(R)$
where the relaxation times $\tau_n$ are independent of the energy $\varepsilon_k$ and, therefore, the coefficient $\eta_{\rm 2DEG}$ vanishes. In this case the spin generated by the current does not depend on the Zeeman splitting and is strictly perpendicular to ${\bm j}$. 
The same result already follows from Eq.~\eqref{S_plus} because, for $F = f''_0$, 
a dependence on $\Omega_{\rm R}$ disappears.
By contrast, anisotropic momentum scattering 
with allowance for energy dependence of the relaxation times
results in the Hanle effect, i.e. an appearance of the spin component parallel to the electric field.
An absence of the Hanle effect for the short-range scattering potential agrees with the analytical result of Ref.~\cite{Wang2010}.
In that work, the calculation for long-range scattering has been performed numerically. Equations~(\ref{s_result}) and~(\ref{eta_2DEG}) present the analytical result for an arbitrary disorder potential~\footnote{
Comparison of Eqs.~(\ref{S_plus}),~(\ref{s_result}) with  Eq.~(9) for the nonequilibrium spin density in Ref. \cite{Wang2012} shows that the true value for the ratio $2C/\Gamma$ is by a factor of 4 smaller than that in Ref. \cite{Wang2012}. As soon as this inaccuracy is removed, the solution of Eq. (9) in Ref. \cite{Wang2012} for the spin ${\bm S}$ becomes unaffected by the out-of-plane magnetization.}.  

In Ref.~\cite{Lazrak2024} a model is considered where there is no scattering between the spin-splitted subbands, but the transport times in the subbands are equal. This approximation gives a reasonable and simple estimate of the conversion tensor components in Eq.~\eqref{CISP_phenom}, but does not describe the Hanle effect in real systems with any rational disorder. In particular, the component $s_\parallel$ does not appear in this model.

\mysection{Spin-orbit coupled graphene}
We consider a heterostructure formed by a graphene layer  coupled to a transition metal dichalcogenide material. Due to the proximity effects, the spin-orbit coupling is introduced to the Dirac fermions in graphene. The Hamiltonian of graphene with spin-orbit coupling and allowance for the valley-Zeeman 
interaction is given by~\cite{Garcia2018,Colloq_RMP2020,Sierra2021}
\begin{multline}
\label{H}
\mathcal H = \mathcal H_0 + \delta \mathcal H, 
  \quad\mathcal H_0=\hbar v_0(\xi  \Sigma_xk_x + \Sigma_y k_y),
\\  \delta \mathcal H= \lambda_{\rm R} (\xi \Sigma_x \sigma_y - \Sigma_y \sigma_x) + \lambda_{\rm VZ}\xi \sigma_z.
\end{multline}
Here $\xi = \pm$ enumerates the valleys in the $\bm K_\xi$ points of the Brillouine zone, $v_0$ is the Dirac fermion velocity, $\lambda_{\rm R,VZ}$ are  values of the Rashba and valley-Zeeman splittings, and two sets of the Pauli matrices $\bm \Sigma$ and $\bm \sigma$ act in the sublattice and spin spaces, respectively.

As well as in the previous 2DEG system, we assume that the Fermi energy is much larger than the spin splittings.
Thus, $\delta \mathcal H$ is a small perturbation, and we can obtain the effective conduction-band Hamiltonian as an average of $\mathcal H$ over the eigenfunctions of $\mathcal H_0$ with positive energy $\varepsilon_k=\hbar v_0 k$, which we choose as $[1,\xi\exp(i\xi\varphi_{\bm k})]^T/\sqrt{2}$~\cite{Golub2024}.
Then we get
\begin{equation}
\mathcal H_c = \varepsilon_k + \lambda_{\rm R} {[\bm k \times \bm \sigma]_z\over k} +\lambda_{\rm VZ}\xi \sigma_z.
\end{equation}
This expression for each valley $\xi =\pm$ is equivalent to the Hamiltonian of a 2D electron system with Rashba and Zeeman splittings Eqs.~\eqref{H_general},~\eqref{Omega} with precession frequencies 
\begin{equation}
\Omega_{\rm R}=-{2\over \hbar}\lambda_{\rm R}, \quad
\Omega_{\rm Z}={2\over \hbar}\xi \lambda_{\rm VZ}.
\end{equation}
The out of plane components are opposite in two valleys.
If graphene forms a heterostructure with a ferromagnet with an out of plane magnetization, then $\Omega_{\rm Z}$ has an additional contribution equal in both valleys.

In the absence of intervalley scattering, 
the spin generation rate for spin-orbit coupled graphene 
is obtained from Eq.~\eqref{G_general}. It is shown in Appendix~\ref{App_spin_gen}, using the relation 
$\qty({1/ \tau_2})' = \qty({12/ \tau_1}-{5/ \tau_2})/ \varepsilon_k$ derived in Appendix~\ref{tau_prime}, that
the spin generation rate has the same form as Eqs.~\eqref{G_par} and~\eqref{G_z}
where the function $F(\varepsilon_k)$ reads 
\begin{equation}
F={(\tau_1 f_0')'\over \tau_2} - 2{f_0'\over \varepsilon_k} \qty({\tau_1 \over \tau_2}-2).
\end{equation}
Substituting this expression into Eq.~\eqref{S_plus}, we calculate the spin polarization by integrating over $\varepsilon_k$ by parts and taking into account that the density of states $g \propto \varepsilon_k$.
For $\Omega_{\rm Z}=0$ we have
\begin{equation}
\label{s_0_Gr}
s_\perp(0) ={\lambda_{\rm R}eEv_0\tau_{\rm tr}\over 2\varepsilon_{\rm F}^2}\qty(3-4{\tau_2\over \tau_{\rm tr}}),
\end{equation}
where $\tau_2$ should be taken at $\varepsilon=\varepsilon_{\rm F}$.
The prefactor here can be expressed via the current density as $\lambda_{\rm R} j/(2v_0\varepsilon_{\rm F}eN)$.
The factor in brackets equals to unity for short-range scattering where $\tau_{\rm tr}=2\tau_2$, and to 5/3 for Coulomb impurity scattering where $\tau_{\rm tr}=3\tau_2$.

In the presence of Zeeman splitting we obtain
Eq.~\eqref{s_result} where 
\begin{equation}
\eta_{\rm Gr}={(\nu-2)\tau_{\rm tr}+4\tau_2\over 3\tau_{\rm tr}-4\tau_2},
\quad
\nu={\dd \ln\tau_{\rm tr}\over \dd \ln\varepsilon_{\rm F}}. 
\end{equation}
For short-range impurity scattering ($\nu=-1$) we have $\eta_{\rm Gr}=-1$,  and for scattering by Coulomb impurities ($\nu=1$),  $\eta_{\rm Gr}=1/5$.

In the limit $\varepsilon_{\rm F} \gg \lambda_{\rm R,VZ}$ considered here, $s_\perp(0)$ matches with the result of Ref.~\cite{Dyrdal2014}. In the regime linear in $\Omega_{\rm Z}$, the component $s_\parallel = \Omega_{\rm Z}\tau_s s_\perp(0)$ coincides with that in Ref.~\cite{Dyrdal2015}.
For an arbitrary relation between the Rashba and Zeeman splittings, the result of Refs.~\cite{Sousa2020,Veneri2022} $s_\perp=s_\perp(0)\Omega_{\rm R}^2/(\Omega_{\rm Z}^2+\Omega_{\rm R}^2)$ for short-range scattering potential follows from Eq.~\eqref{s_result} at $\Omega_{\rm Z}\tau_2\gg 1$ only.

\mysection{Discussion}
First we describe the behavior of CISP in the absence of the Zeeman splitting. 
In 2DEG, according to Eq.~\eqref{eta_2DEG} $s_\perp(0) \propto \tau_{\rm tr}/\varepsilon_{\rm F}$ is independent of the Fermi energy and $\propto 1/\varepsilon_{\rm F}$ for Coulomb and short-range scattering impurities, respectively. In graphene, Eq.~\eqref{s_0_Gr}, in both cases the spin $s_\perp(0) \propto \tau_{\rm tr}/\varepsilon_{\rm F}^2$ is suppressed at large $\varepsilon_{\rm F}$, but it drops as $\propto 1/\varepsilon_{\rm F}$ for Coulomb and as $\propto 1/\varepsilon_{\rm F}^3$ for short-range scattering potentials. Furthermore, for the Coulomb impurities, $s_\perp(0)$ is larger by the factor $5/3$ than for short range scattering at equal Fermi energies and transport relaxation times.

\begin{figure}[t]
	\centering \includegraphics[width=\linewidth]{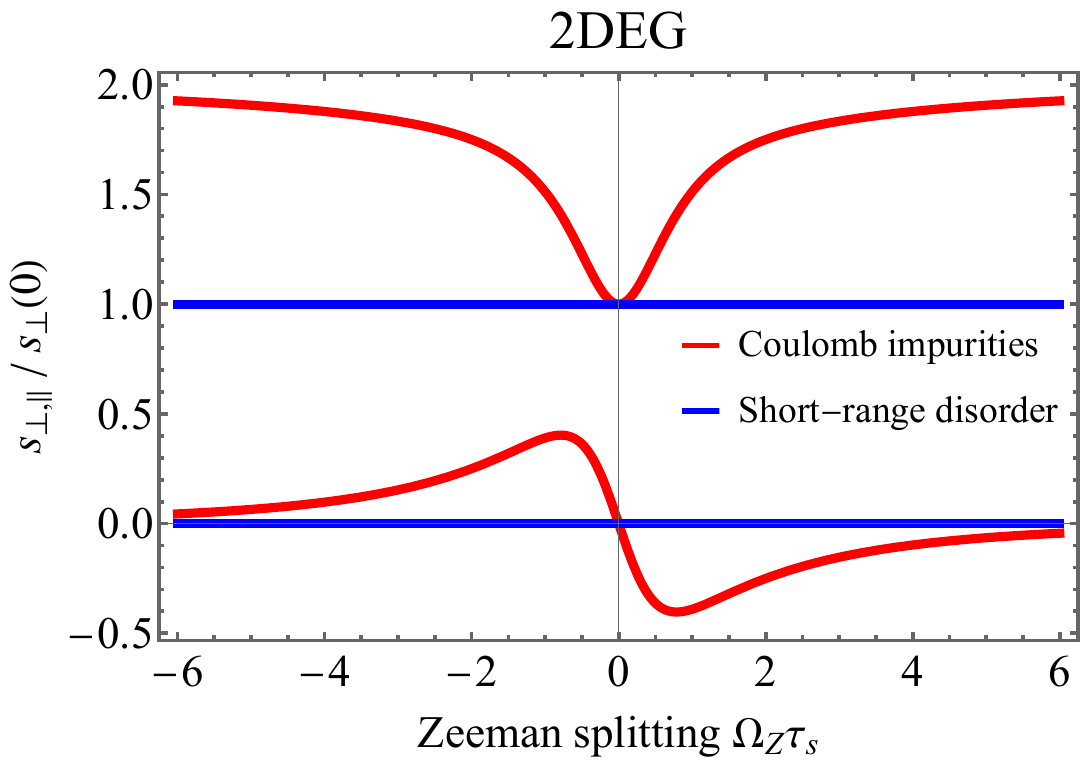} 	
	\caption{Zeeman splitting dependence of CISP in Rashba 2DEG for Coulomb impurity scattering (red) and short-range disorder (blue) at $\Omega_{\rm R}\tau_{\rm tr}=1$. 
Spin components $s_\perp$ are shown in the upper part of the plot, and $s_\parallel=0$ 
%are equal to zero 
at $\Omega_{\rm Z}=0$.
}
	\label{Fig_2DEG}
\end{figure}

Now we turn to the  magnetic field dependence of the CISP.
According to Eq.~\eqref{s_result}, the dependencies of the relative polarizations $s_{\perp}/s_{\perp}(0)$ and $s_{\parallel}/s_{\perp}(0)$ on the dimensionless variable $\Omega_{\rm Z} \tau_s$ is controlled only by one parameter $\Omega_{\rm R} \tau_{\rm tr}$, 
because $\tau_s=2\tau_{\rm tr}/(\Omega_{\rm R} \tau_{\rm tr})^2$.
The out-of-plane magnetization effect on CISP in 2DEG is illustrated
in Fig.~\ref{Fig_2DEG}.
The curves are calculated at $\Omega_{\rm R} \tau_{\rm tr} =1$. 
As stated above, 
for short-range scattering potential the Hanle effect is absent: $s_\perp$ is independent of the Zeeman splitting, and $s_\parallel =0$. On the contrary, for Coulomb impurity scattering the current induced spin 
demonstrates a magnetization induced rotation.
The polarizations $s_{\perp}$ and $s_{\parallel}$ are respectively even and odd functions of $\Omega_{\rm Z}$ in accordance with the parity of real and imaginary parts of the right-hand side of Eq.~\eqref{s_result}.
Since in this case the scattering times $\tau_2=\tau_1/2$ linearly depend on the energy, we have $\eta=1$, and the ratio $s_{\perp}/s_{\perp}(0)$ grows monotonically from 1 to 2 as the product $\Omega_{\rm Z} \tau_s$ varies from 0 to $\infty$.
This is a behavior opposite to that of the conventional Hanle effect observed on optically oriented electrons in semiconductors, e.g.~\cite{Dyakonov2017}.
The $s_\parallel$ component appears owing to the Larmor precession, 
and its absolute value reaches a maximum at some particular value of $\Omega_{\rm Z}$. 
Depending on the relation between the times $\tau_s$ and $\tau_2$  the maximum takes place at $\Omega_{\rm Z} \sim \tau^{-1}_s$ and $\Omega_{\rm Z} \sim (\tau_s \tau_2)^{-1/2}$ for $\tau_2 \ll \tau_s$ and $\tau_2 \gg \tau_s$, respectively.

\begin{figure}[t]
	\centering \includegraphics[width=\linewidth]
	{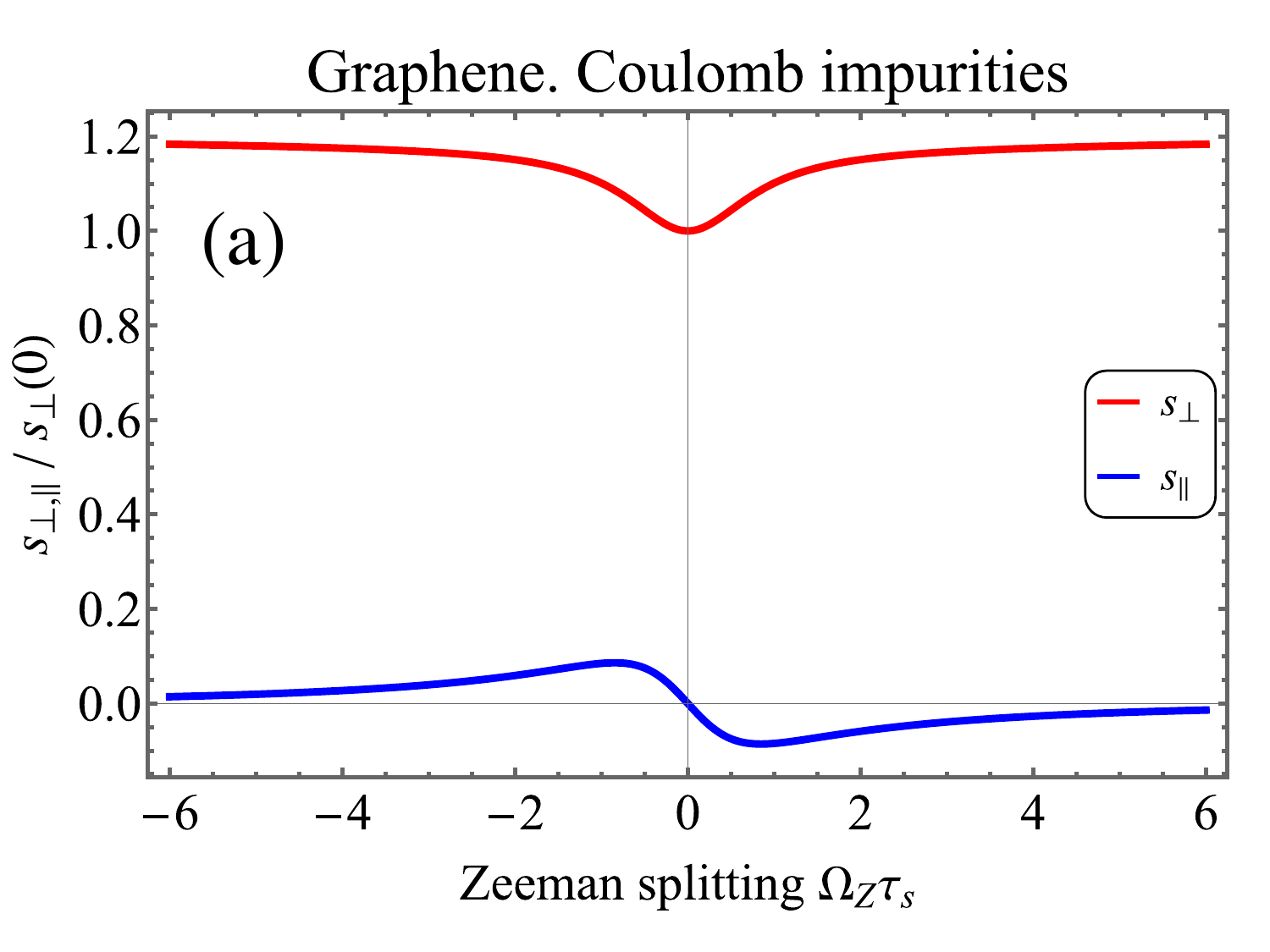}
	\centering \includegraphics[width=\linewidth]
	{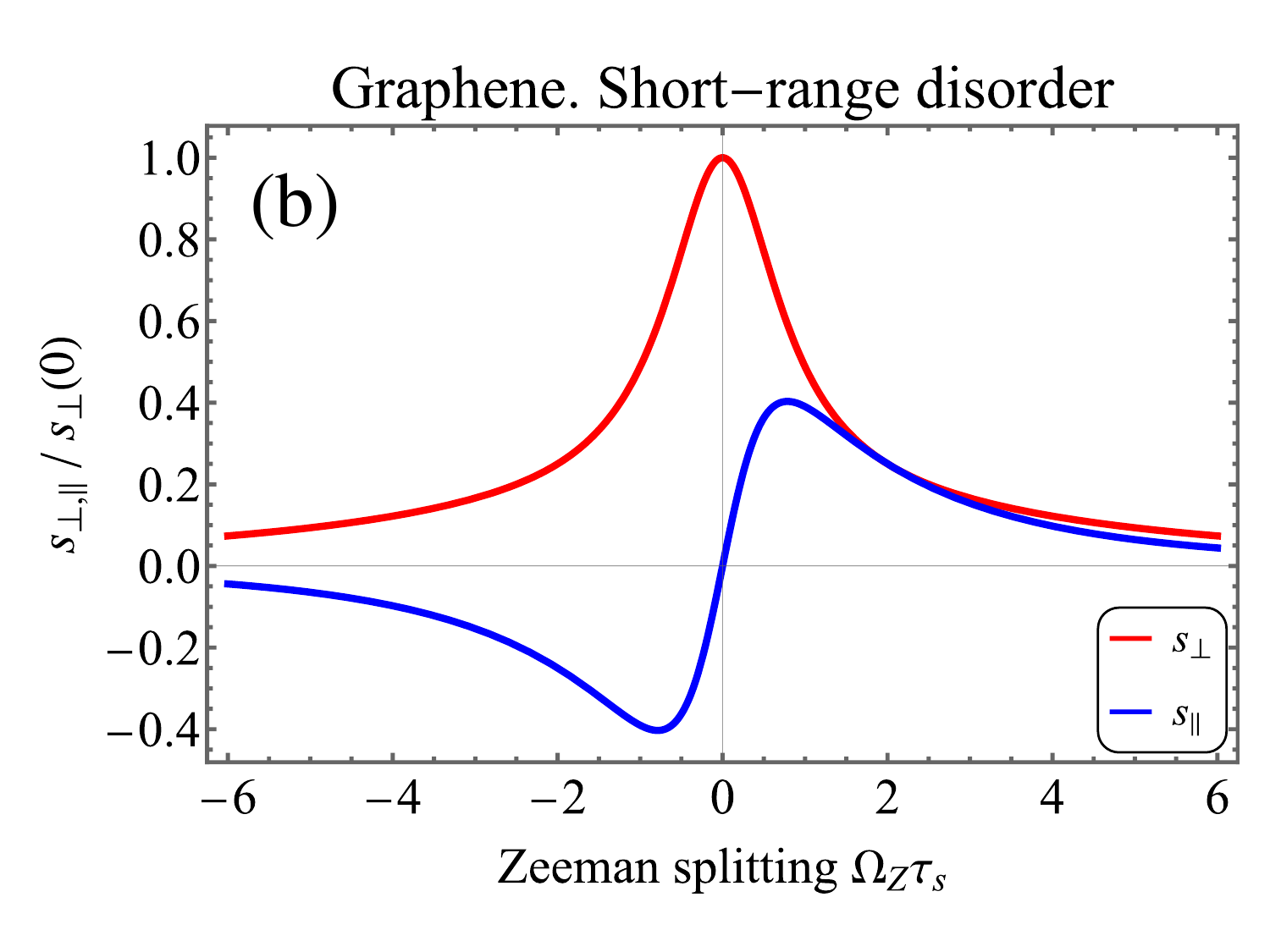}
	\caption{Zeeman splitting dependence of CISP in graphene for Coulomb impurity scattering (a) and short-range disorder (b) at $\Omega_{\rm R}\tau_{\rm tr}=1$. 
	}
	\label{Fig_graphene}
\end{figure}

Dependence of $s_{\parallel,\perp}/s_{\perp}(0)$ in spin-orbit coupled graphene on the Zeeman splitting is shown in Fig.~\ref{Fig_graphene}(a) for the Coulomb scattering potential and Fig.~\ref{Fig_graphene}(b) for short-range scattering. 
One can see that, for a Coulomb potential, the variation of the CISP with increasing value of $\Omega_{\rm Z}$ resembles that in 2DEG. Because of a smaller value of $\eta = 1/5$ the component $s_{\perp}$ increases only by 20\% from its value at $\Omega_{\rm Z}=0$. The magnetization
plays a crucial role in the Hanle effect for short-range scattering in which case $\eta = -1$.  First, the relative perpendicular component $s_{\perp}/s_{\perp}(0)$ monotonously decreases dropping to zero as $\Omega_{\rm Z}$ tends to $\infty$. Second, the relative parallel component $s_{\parallel}/s_{\perp}(0)$ inverses sign and gets comparable with  $s_{\perp}/s_{\perp}(0)$ at $\Omega_{\rm Z} \tau_s \sim 1$.

\mysection{Conclusion}
A kinetic theory of the Hanle effect in CISP is developed for Rashba-coupled 2D systems. Two examples are considered in detail,  electrons in semiconductor heterostructures and Dirac fermions in spin-orbit coupled graphene. It is shown that the Hanle effect in the CISP is very sensitive to the details of electron elastic scattering.
For 2D electrons, CISP is independent of the Zeeman splitting for short-range disorder, while for scattering by long-range Coulomb impurities, the Hanle effect does exist.
For spin-orbit coupled graphene, the spin component perpendicular to the electric field is increased by the out-of-plane magnetization for Coulomb scattering and depressed for short-range disorder. The valley-Zeeman splitting gives rise to the spin component along the electric field opposite in two valleys.
As an outlook, it is important to study the regime of efficient energy relaxation of 2D electrons, frequent inter-valley scattering in spin-orbit coupled graphene, and to generalize the theory to the Dirac systems which covers both the parabolic and linear dispersion as limiting cases and describes the effect in magnetized transition metal dichalcogenide structures.

\mysection{Acknowledgments}
L.E.G. is supported by the Deutsche
Forschungsgemeinschaft (DFG, German Research
Foundation) Project No. Ga501/18 and the Volkswagen Foundation (97738).
E.L.I. acknowledges support from 
the Russian Science Foundation grant No. 22-12-00211.

\newpage

\appendix

\begin{widetext}

\section{Solution of equation for spin density}
\label{App_spin_dyn}

We search the solution of Eq.~(\ref{spin_dyn}) in the form ${\bm S}_{\bm k} = \overline{\bm S} + \delta {\bm S}_1 + \delta {\bm S}_2$, where $\overline{\bm S}$ is independent of the direction of ${\bm k}$,  $\delta {\bm S}_1$ and $\delta {\bm S}_2$ are the 1st and 2nd angular harmonics of ${\bm S}_{\bm k}$. Then the spin dynamics equation can be transferred to
\begin{equation} \label{SSSG}
\left(  \overline{\bm S} + \delta {\bm S}_1 + \delta {\bm S}_2 \right) \times {\bm \Omega}_{\bm k}+ \frac{\delta {\bm S}_1}{\tau_1} + \frac{\delta {\bm S}_2}{\tau_2} - {\bm G}_{\bm k} = 0\:.
\end{equation}
For convenience, we temporarily assume the electric field ${\bm E}$  to be directed along the $x$ axis. Due to the axial symmetry of the Hamiltonian~(\ref{H}) the final result is applicable for arbitrary direction of ${\bm E}$. To solve Eq.~(\ref{SSSG}) we rewrite 
it in the form containing phase factors as follows
\begin{subequations}
\begin{align}
\label{phase1}
&{\bm \Omega}_{\bm k}^{\rm R} = {\bm \Omega}^{\rm R}_k {\rm e}^{{\rm i} \varphi_{\bm k}} + {\bm \Omega}^{{\rm R} *}_k {\rm e}^{-{\rm i} \varphi_{\bm k}}\:, \quad {\bm \Omega}^{\rm R}_k = - \frac{\Omega_{\rm R}}{2} ({\rm i}, 1,0)\:,\\
\label{phase2}
& {\bm G}^{\rm in}_{\bm k} = {\bm G}^{\rm in}_k {\rm e}^{2 {\rm i} \varphi_{\bm k}}  + {\bm G}^{{\rm in}*}_k {\rm e}^{-2{\rm i} \varphi_{\bm k}} \:, \quad {\bm G}^{\rm in}_k = L {\bm \Omega}^{\rm R}_k \:,\\
\label{phase3}
&G_{{\bm k},z} =  G_{k, z}  {\rm e}^{{\rm i} \varphi_{\bm k}}  + G^*_{k, z} {\rm e}^{-{\rm i} \varphi_{\bm k}} \:, \\
\label{phase4}
&\delta {\bm S}_n = {\bm S}_n {\rm e}^{{\rm i}n \varphi_{\bm k}} + {\bm S}^*_n {\rm e}^{-{\rm i}n \varphi_{\bm k}} ~~ (n= 1,2)\:.
\end{align}
\end{subequations}
According to Eqs.~(\ref{G_par}) and (\ref{G_z}), for ${\bm E} \parallel x$  the values of $L$ and $G_{k, z}$ are real. Substituting Eqs.~(\ref{phase1})-(\ref{phase4}) into Eq.~(\ref{SSSG}) we arrive at the equation
$\sum\limits_{l= -3}^3 {\bm D}_l {\rm e}^{{\rm i} l \varphi_{\bm k}}= 0$, 
which gives the following vector equations
\begin{subequations}
\begin{align}
\label{D0}
&{\bm D}_0 = \overline{\bm S} \times ( \Omega_{\rm Z} \hat{\bm z}) + {\bm S}_1 \times  {\bm \Omega}^{{\rm R}*}_k + {\bm S}^*_1 \times  {\bm \Omega}^{\rm R}_k = 0\:,\\
\label{D1}
& {\bm D}_1 =\overline{\bm S} \times  {\bm \Omega}^{\rm R}_k + {\bm S}_1 \times  ( \Omega_{\rm Z} \hat{\bm z}) + {\bm S}_2 \times  {\bm \Omega}^{{\rm R}*}_k
+ \frac{{\bm S}_1 }{\tau_1} - G_{k,z} \hat{\bm z} =0\:, 
\\
\label{D2}
&{\bm D}_2 =  {\bm S}_1 \times  {\bm \Omega}^{\rm R}_k + {\bm S}_2 \times  ( \Omega_{\rm Z} \hat{\bm z}) + \frac{{\bm S}_2 }{\tau_2} -  L {\bm \Omega}^{\rm R}_k =0 \:, \\
\label{D3}
&{\bm D}_3= {\bm S}_2 \times  {\bm \Omega}^{\rm R}_k =0 \:.
\end{align}
\end{subequations}
Since ${\bm D}_{-l} ={\bm D}^*_l$ the harmonics with negative indices do not result in additional equations.

Equation~(\ref{D3}) can be rewritten as ${\bm S}_2 = u {\bm \Omega}^{\rm R}_k$, where the coefficient $u$ is complex and should be found. Then, noting that
${\bm \Omega}^{\rm R}_k \times \hat{\bm z}=-i{\bm \Omega}^{\rm R}_k$, we obtain
 from Eq.~({\ref{D2}}) that ${\bm S}_1 = S_1 \hat{\bm z}$ with $S_1$ comprising both real and imaginary parts. Equation~(\ref{D0}) can be transferred  to
\begin{equation} \label{D0a}
S_1 = - \frac{\Omega_{\rm Z}}{\Omega_{\rm R}} (\overline{S}_y + {\rm i} \overline{S}_x)\:.
\end{equation}
Equation (\ref{D2}) allows us to relate $u$ with $S_1$
\begin{equation} \label{D3b}
u = \frac{S_1 + {\rm i} L}{\Omega_{\rm Z} + {\rm i}/{\tau_2}}\:.
\end{equation}

Substituting $S_1$ and $u$ expressed via $\overline{S}_x, \overline{S}_y$ into Eq.~(\ref{D1}),
using the expressions
\begin{equation} \label{LGz}
L = - \frac{e}{4} F(\varepsilon_{k})\hbar E_x v_k \:,
\quad G_{k,z} = - \frac{e}{4} f''_0 (\varepsilon_{k})\hbar \Omega_{\rm Z} E_x v_k \:,
\end{equation}
where $v_k = \hbar^{-1}\partial_k \varepsilon_k$, 
we get Eq.~\eqref{S_plus} of the main text.

\section{Spin generation rate in spin-orbit coupled graphene}
\label{App_spin_gen}

At $\bm E \parallel y$, Eq.~\eqref{G_general} of the main text yields for the $y$ component of the generation rate 
\begin{multline}
G_{\bm k, y} = -e E_y \lambda_{\rm R}v_0 \partial_{k_y}f_0'{k_x\over \varepsilon_k}
\\
+eE_yv_0\lambda_{\rm R} 
{2\pi \over \hbar}\sum_{\bm k'} \mathcal K_{\bm k'\bm k}\qty[f_0'(\varepsilon_k)\tau_1(\varepsilon_k)\sin{\varphi_{\bm k}}-f_0'(\varepsilon_{k'})\tau_1(\varepsilon_{k'})\sin{\varphi_{\bm k'}}](\cos{\varphi_{\bm k}} - \cos{\varphi_{\bm k'}}) \delta'(\varepsilon_k - \varepsilon_{k'}).
 \end{multline}
Since $\mathcal K_{\bm k'\bm k}$ is an even function of $\theta$, the terms $\propto \sin\theta$ cancel, and we have
\begin{multline}
G_{\bm k, y} = -e E_y \lambda_{\rm R}v_0 {\sin{2\varphi_{\bm k}} \over 2} \qty(f_0'' - {f_0'\over \varepsilon_k})
\\
+eE_yv_0\lambda_{\rm R} {\sin{2\varphi_{\bm k}} \over 2}
{2\pi \over \hbar}
%\sum_{\bm k'} 
\lim_{k'\to k}
\pdv{}{\varepsilon_{k'}} g(\varepsilon_{k'})
\expval{\mathcal K_{\bm k'\bm k}\qty[f_0'(\varepsilon_k)\tau_1(\varepsilon_k)(1-\cos\theta)-f_0'(\varepsilon_{k'})\tau_1(\varepsilon_{k'})(\cos\theta-\cos{2\theta})]}_{\theta}. %\delta'(\varepsilon_k - \varepsilon_{k'}).
 \end{multline}
 Using $\dd g/\dd \varepsilon_{k'} = g(\varepsilon_{k'})/\varepsilon_{k'}$, and
\begin{equation}
\lim_{k'\to k} \pdv{\mathcal K_0(q)}{k'}
= \lim_{k'\to k} \dv{\mathcal K_0(q)}{q}  {k'-k\cos\theta \over q} = \sin({\theta/ 2}) \lim_{k'\to k} \dv{\mathcal K_0(q)}{q},
\end{equation}
while
\begin{equation}
\pdv{}{k}\lim_{k'\to k} \mathcal K_0(q)
= \pdv{\mathcal K_0(2k\sin{\theta/ 2})}{k} 
=2\sin({\theta/ 2})\lim_{k'\to k} \dv{\mathcal K_0(q)}{q},
\end{equation}
we have
\begin{equation}
{2\pi \over \hbar}\lim_{k'\to k}
\pdv{}{\varepsilon_{k'}} g(\varepsilon_{k'})
\expval{\mathcal K_{\bm k'\bm k}(1-\cos{n\theta})}_{\theta}=
{1\over \varepsilon_k\tau_n(\varepsilon_k)} + {1\over 2}\varepsilon_k\pdv{}{\varepsilon_{k}}\qty[{1\over \varepsilon_k\tau_n(\varepsilon_k)}]
= {1\over 2}\qty[\qty({1\over \tau_n})' + {1\over \varepsilon_k\tau_n}],
\end{equation}
where prime means differentiation over $\varepsilon_k$.
Therefore we get
\begin{equation}
G_{\bm k, y} = e E_y \lambda_{\rm R}v_0 {\sin{2\varphi_{\bm k}} \over 2}
\qty[-f_0'' + {f_0'\over \varepsilon_k} +{\tau_1 f_0'\over 2\varepsilon_k}\qty({2\over \tau_1}-{1\over \tau_2}) + {\tau_1 f_0'\over 2}\qty({2\over \tau_1}-{1\over \tau_2})' - (\tau_1 f_0')'\qty({1\over \tau_2}-{1\over \tau_1})],
\end{equation}
which can be simplified to
\begin{equation}
G_{\bm k, y} = e E_y \lambda_{\rm R}v_0 {\sin{2\varphi_{\bm k}} \over 2}
\qty{2{f_0'\over \varepsilon_k}-{\tau_1 f_0'\over 2\varepsilon_k}\qty[{1\over \tau_2}+\varepsilon_k\qty({1\over \tau_2})'] -{ (\tau_1 f_0')'\over \tau_2}} .
\end{equation}
Now we use the relation for elastic momentum relaxation rates of Dirac fermions derived in Appendix~\ref{tau_prime}
\begin{equation}
\label{tau_diff}
\qty({1\over \tau_2})' = {1\over \varepsilon_k}\qty({12\over \tau_1}-{5\over \tau_2}).
\end{equation}

This allows further simplifying the spin generation rate:
\begin{equation}
G_{\bm k, y} = e E_y \lambda_{\rm R}v_0 {\sin{2\varphi_{\bm k}} \over 2}
\qty[2{f_0'\over \varepsilon_k} \qty({\tau_1 \over \tau_2}-2)-{ (\tau_1 f_0')'\over \tau_2}] .
\end{equation}

Analogous calculation of $G_{\bm k, x}$ gives the result which is obtained from $G_{\bm k, y}$ by a substitution $\sin{2\varphi_{\bm k}} \to \cos{2\varphi_{\bm k}}$.
Therefore for the in-plane generation rate vector we can write
\begin{equation}
\label{generation}
\bm G_{\bm k}^{\rm in} = 
eE_y
\qty[2{f_0'\over \varepsilon_k} \qty({\tau_1 \over \tau_2}-2)-{ (\tau_1 f_0')'\over \tau_2}]
{\hbar \over 2} \qty(\bm \Omega_{\bm k}^{\rm R}v_y- \overline{\bm \Omega_{\bm k}^{\rm R}v_y}).
\end{equation}
Note that the prefactor in the generation rate~\eqref{generation} at linear energy dispersion is different from the parabolic-dispersion case, Eq.~\eqref{F_2DEG}.

\section{Derivative of $1/\tau_2$ on energy}
\label{tau_prime}

Here we present the derivation of Eq.~\eqref{tau_diff}.
According to Eqs.~\eqref{tau_n} and~\eqref{Corr} we have
\begin{equation}
{1\over \tau_2(\varepsilon_k)} = {2\pi\over \hbar}g(\varepsilon_k)\expval{ \mathcal K_0(2k\sin{\theta/ 2}){1+\cos\theta\over 2}(1-\cos{2\theta})}_\theta.
\end{equation}
This yields
\begin{equation}
\qty({1\over \tau_2})'  =  {1\over \varepsilon_k\tau_2} + 
{2\pi\over \hbar}g(\varepsilon_k){1\over \hbar v_0} \expval{  \pdv{\mathcal K_0(2k\sin{\theta/ 2})}{k}
 \, {1+\cos{\theta} \over 2} 
(1-\cos{2\theta})}_\theta.
\end{equation}
Using the relation
\begin{equation}
\pdv{\mathcal K_0(2k\sin{\theta/ 2})}{k} = {2\over k}\tan(\theta/2)\pdv{\mathcal K_0(2k\sin{\theta/ 2})}{\theta}
\end{equation}
and integrating by $\theta$ by parts, we obtain
\begin{equation}
\qty({1\over \tau_2})'  =  {1\over \varepsilon_k\tau_2} - 
{2\pi\over \hbar}g(\varepsilon_k){1\over \varepsilon_k} \expval{  {\mathcal K_0(2k\sin{\theta/ 2})}
 \qty[\cos{\theta}(1-\cos{2\theta})+2\sin\theta\sin2\theta]
}_\theta.
\end{equation}
The expression in the square brackets equals to 
$6\cos^2(\theta/2) (2\cos\theta - 1 - \cos{2\theta})$, therefore we get
\begin{equation}
\qty({1\over \tau_2})'  =  {1\over \varepsilon_k\tau_2} - 
{2\pi\over \hbar}g(\varepsilon_k){6\over \varepsilon_k} \expval{  {\mathcal K_0(2k\sin{\theta/ 2})} {1+\cos{\theta} \over 2}  (2\cos\theta - 1 - \cos{2\theta})
}_\theta.
\end{equation}
Finally from definitions~\eqref{tau_n} and~\eqref{Corr} we obtain
\begin{equation}
\qty({1\over \tau_2})'  =  {1\over \varepsilon_k\tau_2} - 
{6\over \varepsilon_k} \qty({1\over \tau_2}-{2\over \tau_1})= {1\over \varepsilon_k}\qty({12\over \tau_1}-{5\over \tau_2}).
\end{equation}
One can check that this relation is valid for both short-range ($\tau_1 = 2\tau_2 \propto 1/\varepsilon$) and Coulomb impurity ($\tau_1 = 3\tau_2 \propto \varepsilon$) scattering.

\end{widetext}

\bibliography{Spin_vdW.bib}
\end{document}